\documentclass[preprint,12pt]{elsarticle}

\usepackage{amssymb}
\usepackage{titlesec}
\usepackage{amsmath}
\usepackage[ruled,vlined]{algorithm2e}
\usepackage{geometry}
\usepackage{epstopdf}
\usepackage{float}
\usepackage{caption}
\usepackage{subcaption}
\usepackage{url}
\journal{arXiv}
\newcommand{\gcxgc}{GC$\times$GC-TOFMS\space}

\begin{document}

\begin{frontmatter}



\title{PARAFAC2$\times$N: Coupled Decomposition of Multi-modal Data with Drift in N Modes}

%%             state={},7

\author[inst1]{Michael D. Sorochan Armstrong}

\author[inst2]{Jesper L\o ve Hinrich}
\affiliation[inst1]{organization={Department of Chemistry, University of Alberta},%Department and Organization
            addressline={11227 Saskatchewan Dr NW}, 
            city={Edmonton},
            postcode={T6G 2G2}, 
            state={Alberta},
            country={Canada}}

\author[inst1]{A. Paulina de la Mata}

\author[inst1]{James J. Harynuk
}

\affiliation[inst2]{organization={Department of Food Science, University of Copenhagen},%Department and Organization
            addressline={Rolighedsvej 26}, 
            city={Copenhagen},
            postcode={DK-1958}, 
            country={Denmark}}

\begin{abstract}
Reliable analysis of comprehensive two-dimensional gas chromatography - time-of-flight mass spectrometry (\gcxgc) data is considered to be a major bottleneck for its widespread application. For multiple samples, \gcxgc data for specific chromatographic regions manifests as a 4\textsuperscript{th} order tensor of \textit{I} mass spectral acquisitions, \textit{J} mass channels, \textit{K} modulations, and \textit{L} samples. Chromatographic drift is common along both the first-dimension (modulations), and along the second-dimension (mass spectral acquisitions), while drift along the mass channel and sample dimensions is for all practical purposes nonexistent. A number of solutions to handling \gcxgc data have been proposed: these involve reshaping the data to make it amenable to either 2\textsuperscript{nd} order decomposition techniques based on Multivariate Curve Resolution (MCR), or 3\textsuperscript{rd} order decomposition techniques such as Parallel Factor Analysis 2 (PARAFAC2). PARAFAC2 has been utilised to model chromatographic drift along one mode, which has enabled its use for robust decomposition of multiple GC-MS experiments. Although extensible, it is not straightforward to implement a PARAFAC2 model that accounts for drift along multiple modes. In this submission, we demonstrate a new approach and a general theory for modelling data with drift along multiple modes, for applications in multidimensional chromatography with multivariate detection

\end{abstract}


\end{frontmatter}


\section{Background}
\label{sec:back}
Multidimensional chromatographic separations are becoming more widespread, thanks in part to advances in modulator technology that have enjoyed considerable interest over the past two decades \cite{bahaghighat2019recent}. The most mature of these technologies is comprehensive two-dimensional gas chromatography (GC$\times$GC) which is frequently hyphenated together with a time-of-flight mass spectrometer\cite{shellie2001application}. \gcxgc is more sensitive and selective than traditional gas chromatography - mass spectrometry, but despite its considerable advantages, and many innovations that have reduced the analysis cost for \gcxgc separations\cite{amaral2019comprehensive}, the technology suffers from challenges surrounding data analysis that hinder its widespread deployment\cite{gorecki_2021}. Currently, few \cite{reichenbach2013reliable,wilde2020automating} software packages offer a transparent and mathematically satisfying way of handling data from untargeted analyses such as those frequently encountered in forensics and metabolomics. Much of the challenge arises due to the fact that chemical components are free to shift independently along the first and second chromatographic modes between runs. This is the major drawback encountered when performing a separation utilising multiple chromatographic modes \cite{li2009comprehensive}, as opposed to tandem mass spectrometric detectors which, due to regular and thorough mass calibrations, do not suffer from mass-to-charge ratio (m/z) drift between runs.

A number of proposals for the analysis of \gcxgc data, based on the well-understood theories of Multivariate Curve Resolution (MCR) \cite{parastar2013solving}, Parallel Factor Analysis (PARAFAC) \cite{hoggard2009toward} and PARAFAC2 \cite{skov2009handling} have been presented in the literature. Models utilising linear rank-deficient solutions have proven to do well to extract meaningful information that is robust against interfering chemical and/or electronic noise\cite{azimi2016multivariate}. The drawback of these techniques is that skilled user intervention is necessary to determine the chemical rank of the data, and identify regions of interest. While PARAFAC2 has shown to be a useful, parsimonious approach to model drifting chromatographic data with multi-channel detectors such as mass spectrometers\cite{bro1999parafac2}, it is limited in that it allows for drift in only one mode.

A number of practical solutions to handling \gcxgc data have been proposed \cite{parsons2015tile}, but these typically lean heavily on the dynamic programming aspect of data analysis\cite{pierce2012review}. Rather than modelling the data, programmatic solutions find, analyze and associate regions of interest across multiple samples and correlate the chemical information for inclusion into a peak table that describes similar chemical characteristics of different samples. This is often done as part of a commercial software solution, or as an additional piece of software designed to work on the peak tables for each individual sample as exported by other software packages \cite{reichenbach2013reliable}. A major issue with dynamic programmatic solutions is that failure of the software at any step can result in misalignment of analytes across multiple samples, or as is more commonly observed, splitting misidentified peaks as separate compounds. In either case, further analysis of imperfect peak tables may lead to erroneous conclusions for untargeted analyses\cite{lu2008comparative}.

There are typically a number of different parameters that require optimisation using the software currently available\cite{weggler2021unique}. Since there is no objective measure for the performance of different data analysis parameters, results that best align with the analysts' expectations are usually assumed to be correct\cite{wilde2020automating}. While the intuition of an experienced analyst is certainly useful, reliance on subjective measures for model performance is far from an ideal solution. Furthermore, for complex mixtures, there is often not an ideal set of parameters that can handle the entire dataset in such a way that matches the expectations of the analyst. For instance, parameters that integrate and align large peaks handily, may miss smaller peaks which fall below integration thresholds. This then obviates the purported advantages of \gcxgc in terms of sensitivity. On the other hand, parameters designed to accurately capture smaller analyte peaks may struggle to handle large, broad, abundant peaks, reporting a single peak as multiple different compounds. In either case, the resultant peak tables pose problems for subsequent data analysis.

%Motivation for this work has been the frustrating and time-consuming chore of curating peak tables for large (i.e. > 200 sample) studies prior to performing multivariate analyses of the data. Curation of peak tables or feature lists currently requires the use of a variety of imperfect commercial and home-built software tools, at the hands of subjective analysts with varying levels of skill and expertise.

This work attempts to remove as many subjective parameters involved in the analysis of \gcxgc\space data as possible, to the end of making pre-processing more efficient, reproducible, and objective. One goal is to avoid integration artefacts such as peak dropout and peak splitting which negatively impact subsequent multivariate analysis. Optimisation of pre-processing parameters is a significant time-sink. In the hands of an experienced analyst, these parameters are typically modified several times in order to find an optimal set that yields a series of peak tables that appear to be of sufficient quality for further work. In the hands of a novice or careless analyst, it is relatively easy to generate a series of peak tables with numerous issues of peak splitting, misalignment, and peak dropout. These will yield meaningless results when further interrogated using multivariate data analysis tools.

Herein, we propose a new mathematical modelling approach for \gcxgc\space that exploits the high degree of redundancy in \gcxgc data sets for a series of samples via a direct decomposition of the 4-way data. This approach is based on the flexible coupling method for 3-way PARAFAC2, with an additional coupling constraint that restricts the descent of the extracted mass spectra calculated from models that describe the first- and second-dimension retention drifts. While extremely useful for analyzing \gcxgc data, this technique offers a general theory for modelling multidimensional chromatographic data with drift in $N$ modes, and may also be extensible to hyperspectral imaging datasets. Much like PARAFAC2, the proposed algorithm we are calling PARAFAC2$\times$2, requires only the number of components and a region of interest in order to work. This greatly simplifies the task of analysing \gcxgc data, removing the long lists of parameters to be optimised and subjectivity in data analysis.

\subsection{\gcxgc Data Structure}
A univariate detector such as a flame-induction detector (FID) performs a series of regular measurements at regular intervals as chemical components enter the detector from the GC column. Certain regions of a single chromatogram with one dimension of separation can be excised to analyze the chromatographic peak in question for quantitative purposes. An excised region is a vector of length $I$, where $I$ is the number of acquisitions in the region of interest. The relative abundance of peaks in this region can be obtained by the euclidean norm of the vector, assuming no interfering analytes are present within the region, or by a non-linear, parametric fit of several idealised Gaussian or modified Gaussian \cite{asher2009comparison} functions to deconvolve the signals of chemical interference.

When the separation is coupled to a multivariate detector, such as a mass-spectrometer or a vacuum UV detector, the number of different variables encompassed by the detector can span an additional mode, denoted as $J$. An excised chromatographic region encompasses the number of acquisitions, $I$ by the number of individual ``detectors" (e.g. mass-to-charge ratios, or m/z, for mass spectral detectors), $J$. Closely co-eluting factors can be deconvolved using multivariate methods such as MCR or ICA, both of which decompose the resultant $I\times J$ matrix.

A single \gcxgc chromatogram presents itself as a 3\textsuperscript{rd} order tensor, while a series of \gcxgc chromatograms presents itself as a 4\textsuperscript{th} order tensor comprising $I \times J \times K \times L$ modes of mass spectral acquisitions, mass-to-charge ratios (mass channels), modulations, and samples. The multidimensional separation is generated by capturing fractions of effluent from the first dimension and injecting them at regular intervals onto the second-dimension column via the modulator. The action of the modulator creates slices of second-order information along the first chromatographic dimension, such that for an individual \gcxgc sample the data is a 3\textsuperscript{rd} order tensor. The $L^{th}$ mode describes multiple samples extracted from either the same region of the chromatogram, or entire chromatograms depending on what is being considered.

\subsection{PARAFAC Modelling of \gcxgc Data}
A PARAFAC model can be constructed that describes a 4\textsuperscript{th}- order tensor, $\mathcal{X} \in \mathbb{R}^{I \times J\times K\times L}$, using the Khatri-Rao (KR) product (Appendix \ref{app:KR}) \cite{kolda2009tensor}:

\begin{equation}\label{parafacDot}
    X = F_2(D_l\odot F_1\odot A)^T
\end{equation}

Where $X \in \mathbb{R}^{I\times J*K*L}$ rearranged from $\mathcal{X}$, $F_2$ is as $I\times R$ matrix, $A$ is a $J\times R$ matrix, $F_1$ is a $K\times R$ matrix, and $D_l$ is an $L \times R$ matrix that corresponds to the characteristics of the data mentioned previously in addition to the $R$ chemical factors that best represent the characteristics of the data being analysed. 

A trilinear decomposition of the unfolded tensor $\mathcal{X} \in \mathbb{R}^{I*K \times J \times L}$ of the data can be made, observing that the $KR$ product of the second-dimension elution profiles ($I\times R$) and the modulation matrices ($K \times R$) are equal to a single unfolded retention mode, of dimension $I*K \times R$:

\begin{equation}\label{modes}
    X_l = (F_2\odot F_1)D_LA^T
\end{equation}

$X_l \in \mathbb{R}^{I*K \times J \times L}$ is structurally similar to the trilinear PARAFAC1 model as described by Kiers and Bro, substituting $F$ for $(F_2\odot F_1)$:

\begin{equation}\label{parafac}
    X_k = FD_kA^T
\end{equation}

$X_k \in \mathbb{R}^{I\times J\times K}$ is a series of matrices. This notation is used to keep the notation consistent with the original direct fitting algorithm for PARAFAC2 \cite{bro1999parafac2}.

It is relatively straightforward to frame the problem as a PARAFAC2 model with multiple samples' first- and second-dimensions unfolded as one, with $L$ samples:

\begin{equation}\label{parafac2}
    X_l = (F_2\odot F_1)_lD_lA^T
\end{equation}

Using the direct-fitting method, there are $l$ unique, orthogonal peak profiles of $I*K\times R$, $P_l$, and a non-singular $R\times R$ matrix, $F$:

\begin{equation}\label{parafac2orth}
    X_l = P_lFD_lA^T
\end{equation}

In addition to unfolding the $\mathcal{X} \in \mathbb{R}^{I \times K \times J \times L}$ tensor along the first and second retention modes to effect an $X_l \in \mathbb{R}^{I*K\times J\times L}$ 3\textsuperscript{rd} order tensor, tensors of similar orders can be made by ``stacking" second-dimension retention profiles for an $X_{kl} \in \mathbb{R}^{I\times J \times K*L}$, or the first-dimension retention times' equivalent as: $X_{il} \in \mathbb{R}^{K\times J\times I*L}$. In all cases, it is possible to construct a PARAFAC2 model on the resultant trilinear data. In the first case, the $X \in \mathbb{R}^{I*K\times J\times L}$ appears to avoid the problem of drift in two modes, by artificially reducing the problem to drift along one combined retention mode. This method appears to have an additional benefit, wherein the quantities of each component are solved for directly. It is not possible to solve for the relative expression of each component, per sample directly using the unfolded data in the other two cases. 

\subsection{PARAFAC2 modelling for 4-way data unfolded as: $X_l \in \mathbb{R}^{I*K\times J \times L}$}

\begin{figure}[ht]
    \centering
    \includegraphics[width=\linewidth]{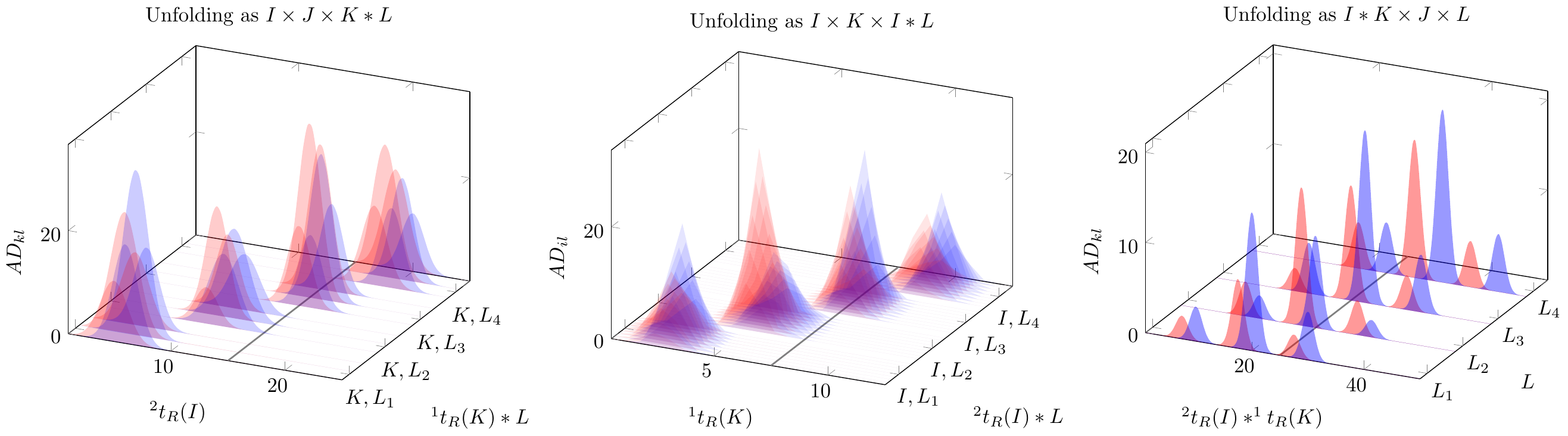}
    \caption{Three possible methods for unfolding GC$\times$GC-TOFMS data for subsequent PARAFAC2 modelling. PARAFAC2$\times$2 uses the two leftmost methods as intermediaries for the decomposition of an entire 4\textsuperscript{th}-order tensor with drift in two modes.}
    \label{fig:unfold}
\end{figure}

A PARAFAC2 model for data unfolded as: $X_l \in \mathbb{R}^{I*K\times J\times L}$ may appear to account for drift in two modes (Figure \ref{fig:unfold}); however, there are practical limitations to the PARAFAC2 model. PARAFAC2 can only account for small, independent variations in retention time drift, based on the assumption that the inner-product matrices: $F_l^TF_l$ (unfolded scores matrices from Equation \ref{parafac2}) are consistent across all samples. In the method for direct fitting of PARAFAC2, $F_l$ is defined as $P_lF$, where $P_l$ are the orthonormal scores matrices that are free to vary across each sample, calculated as:

\begin{equation}
    P_l = X_lAD_lF^T(FD_lA^TX_l^TX_lAD_lF)^{-1/2}
\end{equation}

Through the singular value decomposition of:

\begin{equation}
    FD_lA^TX_l^T = U_k\Sigma_lV_l^T
\end{equation}

\begin{equation}
    P_l = V_lU_l^T
\end{equation}

Because $F_l^TF_l$ is calculated as $F^TP_l^TP_lF$, and because $P_l$ is orthonormal such that $P_l^TP_l = I_R \in \forall l$, $F^TP_l^TP_lF = F^TF$. Consequently, $F$ itself is assumed to be constant across all samples, and is calculated as a sum even for those samples where it differs.

$F$ is calculated from the PARAFAC model of $P_l^TX_l$ which minimises:

\begin{equation}
    P_l^TX_l = argmin_{F,D_l,A}||P_l^TX_l - FD_lA^T||^2_F
\end{equation}

Which can also be described via:

\begin{equation} \label{calculateF}
    F = \sum_{l=1}^{L} P_l^TX_lAD_l(D_lA^TAD_l)^{-1}
\end{equation}

The $P_l^TX_l$ term presents mass spectral, or second-mode loading information that is also proportional to the relative abundance of each chemical factor. Across $L$ samples, this information will be relatively consistent, as long as $P_l$ is describing the latent chemical phenomena in the same way. For small variations in retention time, this is not usually a problem. Small retention time drifts of each component relative to each other may not be significant, and small modelling errors are summed via the calculation of $F$ in Equation \ref{calculateF}. However for unfolded data, small drifts across the first retention mode are in practice large drifts across the combined first-and second-dimension modes. This problem can be mitigated using the flexible coupling approach for non-negative PARAFAC2 by Cohen and Bro, which does not rely on the intermediate calculation of orthogonal peak profiles, and permits modelling on more substantial retention drift relative to the different chemical factors thanks to softer constraints on modelling the data.

\subsection{A Flexible Coupling Approach for Non-negative PARAFAC2}
Cohen and Bro \cite{cohen2018nonnegative} proposed a flexible coupling method for modelling non-negative scores along the mode that is allowed to vary in the PARAFAC2 model. Using this technique, on a 3\textsuperscript{rd} order tensor, $X_k \in \mathbb{R}^{I\times J\times K}$ the non-negatives scores, $B_k$, are calculated as the minimisation of:

\begin{equation} \label{nnparafac}
    X_k = argmin_{B_k,D_k,A,P_k,B^*}\sum_{k=1}^K||X_k - B_kD_kA^T||_F^2 + \mu_k||B_k - P_kB^*||_F^2
\end{equation}

Where non-negativity can be enforced for any term with any non-negative least squares solver. The $P_k$ and $B^*$ terms in Equation \ref{nnparafac} are the orthonormal scores of $B_k$ via SVD, and an $R\times R$ latent coupling factor, which together minimise the second term proportional to the coupling factor, $\mu_k$. The flexible coupling approach for calculating non-negative PARAFAC2 can be implemented using the ALS algorithm although the numerical stability depends on an appropriate estimate for the coupling constants, $\mu_k$. As the solution approaches a minimum, it is reasonable to increase the coupling constant to tighten restriction on the coupled terms.

\begin{equation}
    \mu_k^1 = 10^{-SNR/10}\frac{||X_k - B_k^1D_k^1A^{1T}||^2_F}{||B_k^1 - P_k^1B^{1*}||^2_F}
\end{equation}

After the first iteration of the algorithm, where $SNR$ is the estimated Signal-to-Noise Ratio for each chromatographic slice. A convenient estimate of the $SNR$ can be used by calculating the ratio of the first singular value to the second singular value for column-centred data. The first singular value can be thought of as the distance along the axis of greatest variance within the data from zero, and the second singular value as an estimate for the noise.

It is possible to solve for the unfolded scores, and sample-wise relative abundances using the flexible coupling approach for data unfolded along one retention mode as $X_l \in \mathbb{R}^{I*K\times J \times L}$

\begin{equation}
    X_l = argmin_{B_l,D_l,A,P_l,B^*}\sum_{l=1}^L||X_l - B_lD_lA^T||_F^2 + \mu_l||B_l - P_lB^*||_F^2
\end{equation}

This helps to avoid the issue of inconsistent cross-products that limit the accuracy of the direct-fitting model where there are significant drifts of the chemical components relative to one another. However the scores matrix, $B_l$, is of a relatively high dimensionality at $I*K$ unique indices. This introduces a high number of degrees of freedom, at the expense of the high number of replicates it is possible to achieve by unpacking \gcxgc data in a different fashion. It is well known that PARAFAC models benefit from relatively high numbers of replicates, playing upon the rotational determinacy of PARAFAC models versus analogous matrix decomposition techniques.

\section{PARAFAC2 modelling of 4-way data unfolded as: $X_{kl} \in \mathbb{R}^{I\times J\times K*L}$ or $X_{il} \in \mathbb{R}^{K\times J \times I*L}$}
A \gcxgc dataset comprised of multiple samples can also be unfolded into two third-order tensors as either consecutive slabs of second-dimension retention slices, or first-dimension retention slices ( $X_{kl} \in \mathbb{R}^{I\times J \times K*L} = \mathcal{X}_{:,:,k,l}$ or $X_{il} \in \mathbb{R}^{K \times J \times I*L} = \mathcal{X}_{i,:,:,l}$). The matrices $B_{kl}$ and $B_{il}$ are score matrices for the $hl^{th}$ unfolding of the tensor $\mathcal{X}_{ijkl}$, where $h$ is one of $k$ or $i$. The same notation applies to all of the other matrices with similar designations. 

\begin{equation}\label{profkl}
    X_{kl} = argmin\sum_{kl = 1}^{KL}||X_{kl} - B_{kl}D_{kl}A^T||_F^2 + \mu_{kl}||B_{kl} - P_{kl}B_{kl}^{*}||_F^2
\end{equation}

\begin{equation}\label{profil}
    X_{il} = argmin\sum_{il = 1}^{IL}||X_{il} - B_{il}D_{il}A^T||_F^2 + \mu_{il}||B_{il} - P_{il}B_{il}^{*}||_F^2
\end{equation}

The advantage with these two methods for unfolding, is that there are either $K*L$ or $I*L$ numbers of replicates. The problem of inconsistent cross-product matrices is not eliminated however, as in either case peaks invariably disappear and reappear for properly selected regions of interest. As mentioned earlier, the problem of inconsistent cross-product matrices is mitigated through the use of the flexible coupling approach for non-negative PARAFAC2.

%Fix this - B is doubly defined
The advantage of unfolding the data as a series of second- or first-dimension elution profiles is that it exploits the high degree of redundancy of \gcxgc data. This plays into the advantages of PARAFAC over second-order modelling. However, further manipulation of the resultant scores is required in order to solve for the sample-wise relative abundances. This is simple to do - the scores of either $B_{kl}$ or $B_{il}$ can be unfolded for each sample as an $I*K\times R$ matrix and the data matrix itself unfolded as an $I*K\times R$ matrix. The sample-wise abundances can be solved for in the least-squares sense, where by unfolding $B \in \mathbb{R}^{I*K\times R \times L}$, it is possible to solve for each $l^{th}$ slice of the tensor $D_l$ with the similarly unfolded tensor $X_l \in \mathbb{R}^{I * K\times J \times L}$:

\begin{equation}
    D_l = (B^TB^{-1})BX_lA(A^TA)^{-1}
\end{equation}

Where $A$ is the $J\times R$ matrix of the extracted mass spectra, common to the entire dataset.

There are two further advantages of unfolding the data as first- or second-dimension retention slices: unimodality constraints can be applied using an appropriate least-squares solving algorithm \cite{bro1998least}, and modelling one sample is more extensible to further samples, since there are replicates even within a single chromatographic sample.  Applying the calculated model to new data would make it easy to quantify analytes of interest across different analysis conditions, assuming their location in the chromatogram is known.

\section{The PARAFAC2$\times$2 algorithm}
While it is clear that a \gcxgc dataset can be decomposed in a manner that preserves the high degree of redundancy, the question remains as to which retention mode to model. In theory, the retention mode to model should have the highest resolution between closely co-eluting chemical factors. However this information would be difficult to predict, without first calculating the model itself. 

In order to select an appropriate method for unfolding the data, and with an eye towards creating the most general solution possible for the deconvolution and quantification of \gcxgc features, we propose a method that models both models simultaneously (using Equations \ref{profkl} and \ref{profil}), and at convergence averages the elution scores and corresponding mass spectra for each of the modelled retention modes to solve for the sample-wise loadings. The mass spectra for each model should be allowed to vary slightly, since while the data is the same for both models, the descent to an optimum may differ along the two retention modes. The unified model can be described as the minimisation of the following expression:

\begin{multline}\label{fullexp}
    \mathcal{X}_{ijkl} = argmin_{B_{kl,il},D_{kl,il},A_{kl,il},P_{kl,il}B_{kl,il}^*}\sum_{kl = 1}^{KL}||X_{kl} - B_{kl}D_{kl}A^T||_F^2 + \mu_{kl}||B_{kl} - P_{kl}B_{kl}^{*T}||_F^2 + \\ \sum_{il = 1}^{IL}||X_{il} - B_{il}D_{il}A^T||_F^2 + \mu_{il}||B_{il} - P_{il}B_{il}^{*T}||_F^2 + \\ \mu_A||A_{kl} - A_{il}||_F^2
\end{multline}

The descent that minimises the sum of residual squares for each model informs the learning of the other via the mass spectral coupling constant, $\mu_A$. As long as an appropriate value for $\mu_A$ is selected, this method is readily able to converge to a usable solution in relatively few iterations. However if this coupling constant is too small the iterations may begin to diverge, and if it is too large then the model may limit the descent of the mass spectra, and converge to a sub-optimal solution. It is also important to be cognisant of risk of converging to sub-optimal solutions, depending on the initialisation of the algorithm. For this reason, as is commonly done with PARAFAC2, 10 random initialisations were utilised and the sum of residual squares was measured after 80 iterations. The model with the lowest sum of residual squares is selected as the ``best" initialisation, and is allowed to continue to convergence.

The description of the algorithm in its current implementation follows. For each least-squares step, constraints such as non-negativity or unimodality can be applied depending on what is deemed appropriate for the data.
%\begin{changemargin}{+3cm}{+3cm}
\newgeometry{margin = 2.5cm}
\begin{algorithm}[H]\label{parafac2x2}
\SetAlgoLined
\KwResult{$\mathcal{F}$,$D_l$,$A$ = PARAFAC2$\times$2($\mathcal{X}$,$R$)\\
$\mathcal{F} \in \mathbb{R}^{I\times R\times K\times L}$, $D_l \in \mathbb{R}^{R\times R \times L}$, $A \in \mathbb{R}^{J\times R}$, and $\mathcal{X} \in \mathbb{R}^{I\times J\times K \times L}$
}
 initialization: $B^0_{kl} = rand(I,R,K*L)$,\space $B^0_{il} = rand(K,R,I*L)$\\ $A^0_{il} = A^0_{kl} = rand(J,R)$,\space $B^{*0}_{kl} = B^{*0}_{il} = rand(R,R)$\\ $D^0_{kl} = I_R, \forall {kl} \in [1,K*L]$,\space $D^0_{il} = I_R, \forall {il} \in [1,I*L]$\\ $\mu_{kl}^0 = \frac{||B_{kl}D_{kl}A^T||_F^2}{||B_{kl}||_F^2}$,\space
 $\mu_{il}^0 = \frac{||B_{il}D_{il}A^T||_F^2}{||B_{il}||_F^2}$
 $\mu^0_A = 10^{\omega}\frac{||X_{kl} - B^0_{kl}D^0_{kl}A^{0T}||_F^2 + ||X_{il} - B^0_{il}D^0_{il}A^{0T}||_F^2}{||A_{kl}||_F^2}$\\
 $iter = 1$\\
 \While{$\frac{\sigma_{old} - \sigma_{new}}{\sigma_{old}} > \epsilon\sigma_{old}$}{
 $\sigma_{old} = \sigma_{new}$\\
 \For{$h,H \in k,i \And I,K $}{
  \For{$\forall{hl} \in [1,H*L]$}
  {$[U,\Sigma,V] = SVD(B_{hl}*B^{*}_{hl},R)$\\
  $P_{hl} = UV^T$\\}
  $B^{*}_{hl} = \sum_{hl = 1}^{H*L}\mu_{hl}P_{hl}^TB_{hl}$\\
  $A_{hl} = \sum_{hl = 1}^{H*L}\left(\frac{\mu_AA_{hl} + X_{hl}^TB_{hl}D_{hl}}{D_{hl}B_{hl}^TB_{hl}D_{hl} + \mu_AI_R} \right)$ \%\% See Appendix A \\
  $B_{hl} = \frac{X_{hl}A_{hl}D_{hl} + \mu_{hl}P_{hl}B_{hl}^*}{D_{hl}A_{hl}^TA_{hl}D_{hl} + \mu_{kl}I_R}$ \%\% See Appendix B \\
  \For{$\forall{hl} \in [1,H*L]$}
  {$D_{hl} = \frac{B_{hl}^TX_{hl}A_{hl}}{(B_{hl}^TB_{hl})(A_{hl}^TA_{hl})}$}
  
  \eIf{$iter = 1$}{
   \For{$\forall{hl} \in [1,H*L]$}
   {$\Sigma = SVD(X_{kl},2)$ \\
   $SNR \approx \Sigma_1/\Sigma_2$\\
   $\mu_{hl} = 10^{-SNR/10}\frac{||X_{hl} - B_{hl}D_{hl}A_{hl}^T||_F^2}{||B_{hl} - P_{hl}B_{hl}^*||_F^2}$
   }
   }{
   \If{$iter < 10$}
   {\For{$\forall{hl} \in [1,H*L]$}{$\mu_{hl} = \mu_{hl}*1.05$} 
   }
  }
  }
  $\sigma_{new} = \sum_{h = k}^i||X_{hl} - B_{hl}D_{hl}A_{hl}^T||_F^2 + ||B_{hl} - P_{hl}B_{hl}^*||_F^2 + ||A_{il} - A_{kl}||_F^2$\\
  $iter = iter + 1$
  }
  $\mathcal{F} = B_{kl}D_{kl} + B_{il}D_{il}$\space $\forall {l} \in [1,L]$\\
  $D_l = \frac{F_{I*K\times R}^TX_{I*K\times J \times L}A_{J\times R}^T}{(F_{I*K\times R}^TF_{I*K\times R})(A_{J\times R}^TA_{J\times R})}$\space $\forall {l} \in [1,L]$\\
  $A = A_{kl} + A_{il}$

  \caption{Coupled PARAFAC2$\times$2 ALS}
  \end{algorithm}
  %\end{changemargin}
  \restoregeometry

$B_{kl},B_{il}, A_{kl}, A_{il}, B_{kl}^*,B_{il}^*$ are all explicitly normalised column-wise in Algorithm \ref{parafac2x2}, following the form: $||A_{:,r}||_2 = 1 \space \forall_r \in [1,R]$. $P_{kl}$, and $P_{il}$ are implicitly normalised through their determination via SVD. $\mathcal{F}$ is normalised explicitly as $||\mathcal{F}_{:,r,:,l}||_F = 1 \space \forall_r \in [1,R] \And \forall_l \in [1,L]$, which is equivalent to $||B_{:,r}||_2 = 1 \space \forall_r \in [1,R] \And \forall_l \in [1,L]$, where $B \in \mathbb{R}^{I*K \times R \times L}$. For the initial mass spectral coupling constant, $\mu_A$, an additional exponential term, $\omega$, is added to control the initial descent of the two modes with respect to each other. That is, typically a value of 2 or 3 is used so that the two calculated mass spectra do not diverge at the outset.

\subsection{Analysis of Synthetic Data using PARAFAC$2\times$2}
Synthetic data, mimicking replicate samples of \gcxgc data was generated in in MATLAB\textsuperscript{\textregistered} 2021a to evaluate the performance of the algorithm. Random independent drift in both the first- and second-dimension retention modes across three samples was chosen. For the first-dimension retention time, each peak was allowed to vary $\pm$1.5 modulations, and for the second-dimension retention time each peak apex was allowed to vary randomly $\pm$25 acquisitions. Relatively few replicate samples were chosen for this data set to make the results easier to display, and to demonstrate this algorithm's utility despite handling relatively few samples. Synthetic mass spectra were generated from random distributions of 45 ``peaks" representing isotopic mass distributions, and each spectrum was normalised to its Euclidean norm. Nominally the SNR was set to 500, which is a fair representation of \gcxgc data. White noise was added across every acquisition, modulation, and mass channel relative to the maximum score value out of all components, per unit of SNR. An additional offset of 6 times the maximum score value out of all components per unit of SNR was added to ensure that all data was positive. The distribution for each peak along the first-dimension retention time was set at 1.5 modulations, and along the second-dimension retention time 20 acquisitions. The magnitude of the data was multiplied by a factor of $10^{4}$ to simulate ion counts typically encountered during \gcxgc experiments. The synthetic data was then inspected to ensure a visual similarity to real data, by examining the retention profiles across all channels. The code used to generate the synthetic data is available online at \url{https://github.com/mdarmstr/parafac2x2}. 

Hyperparameters and convergence criteria for the PARAFAC2$\times$2 algorithm as applied for the synthetic experiments were as follows: $\epsilon$ was set at $2.5\times 10^{-6}$, and the number of iterations during which the coupling constants $\mu_{kl}$, $\mu_{il}$, and $\mu_A$ the coupling constants increased was 10. The non-negative solver used was the Fast Combinatorial Non-negative Least Squares algorithm \cite{van2004fast} for the mass spectral and elution modes. However in practice a non-negative least squares solver does not appear to be necessary, as the mass spectral coupling term appears to constrain the solution to be positive. A constant of 3 was used as the value for $\omega$. In both cases, the algorithm converged to a solution in fewer than 30 iterations, which took less than 30 seconds in total. Initialisations were not replicated for these data.

Because of the way the data is generated, the expected peak intensities depend on the maximum score of each of the input factors. This maximum score varies, depending on the first-dimension retention times (i.e. a Gaussian along the first dimension may be modulated close to its apex in one sample, and further away from its apex in another sample), but the recovered abundances are in relatively good agreement with our expectations despite this limitation in precision.

The cosine correlation coefficient ($\cos(\theta)$) was used to measure the agreement between the calculated and synthetic scores and loadings ($\nu_{SYN}$, and $\nu_{OBS}$) using the unfolded GC$\times$GC scores and mass spectral loadings. This was calculated as the inner product of each pair of matrices, with each column normalised to its Euclidean norm such that for each pair of entries:

\begin{equation}
    \cos(\theta) = \frac{\nu_{SYN}\cdot\nu_{OBS}}{||\nu_{SYN}||\cdot||\nu_{OBS}||}
\end{equation}

The percent variance explained was calculated using the formula from Bro et al. \cite{bro1999parafac2}

\begin{equation}
    \% VAR = 100\times\left(1-\frac{\sum_{i=1}^{I}\sum_{j=1}^{J}\sum_{k=1}^{K}\sum_{l=1}^L\left(X_{ijkl}-FD_lA^T\right)^2}{\sum_{i=1}^{I}\sum_{j=1}^{J}\sum_{k=1}^{K}\sum_{l=1}^L\left(X_{ijkl}\right)^2}\right)
\end{equation}

Shown below are the results of analysing a synthetic two-component elution. For ease of comparison, an extra component was not used to model the noise.

\begin{figure}[H]
    \centering
    \includegraphics[width=\textwidth]{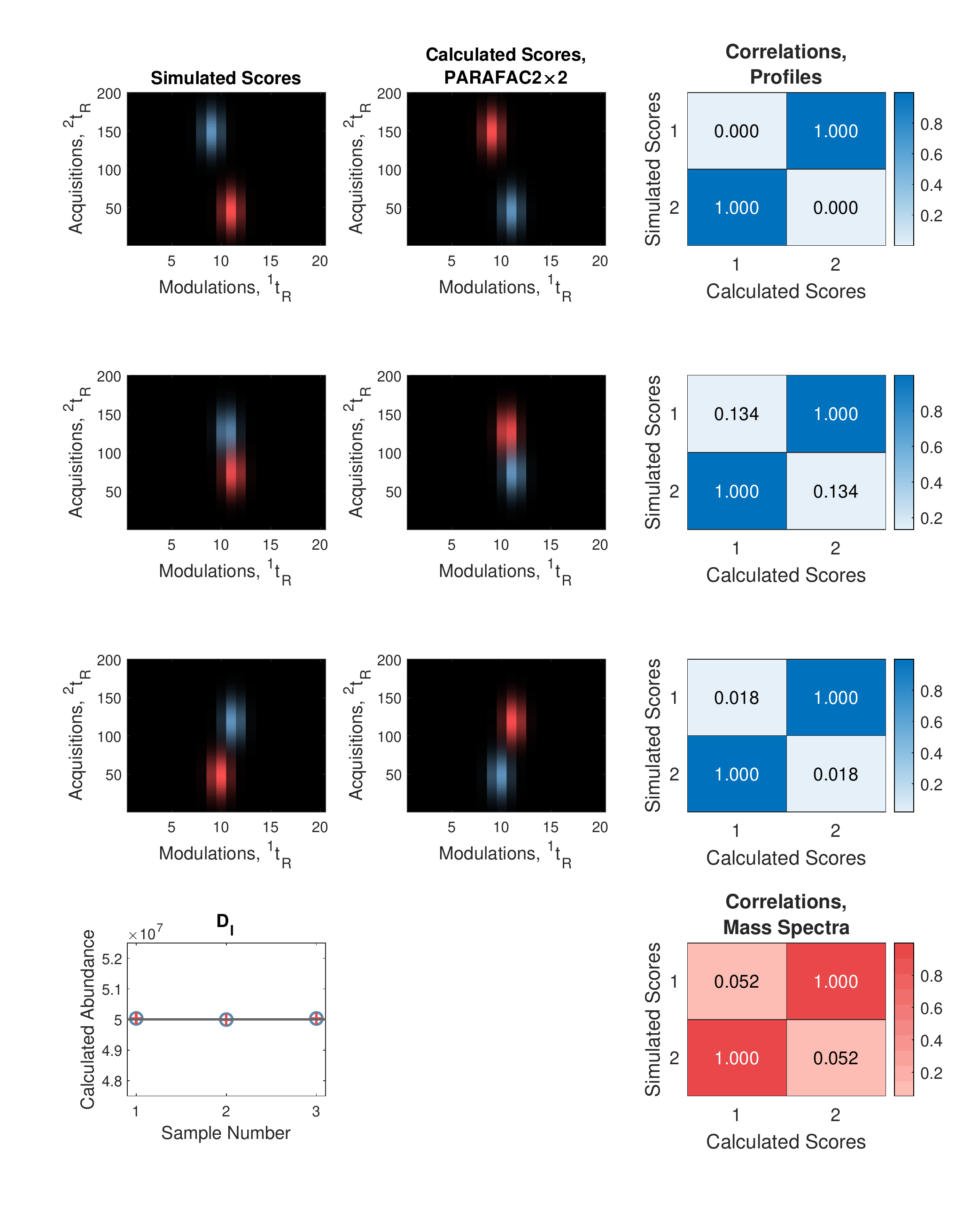}
    \caption{A simulated two-component model with a nominal SNR of 500. The percent variance explained using this model was 99.9959\%. The calculated model demonstrates almost perfect agreement with the synthetic data.}
    \label{fig:2comp}
\end{figure}

\newpage
\begin{figure}[H]
    \centering
    \includegraphics[width=\textwidth]{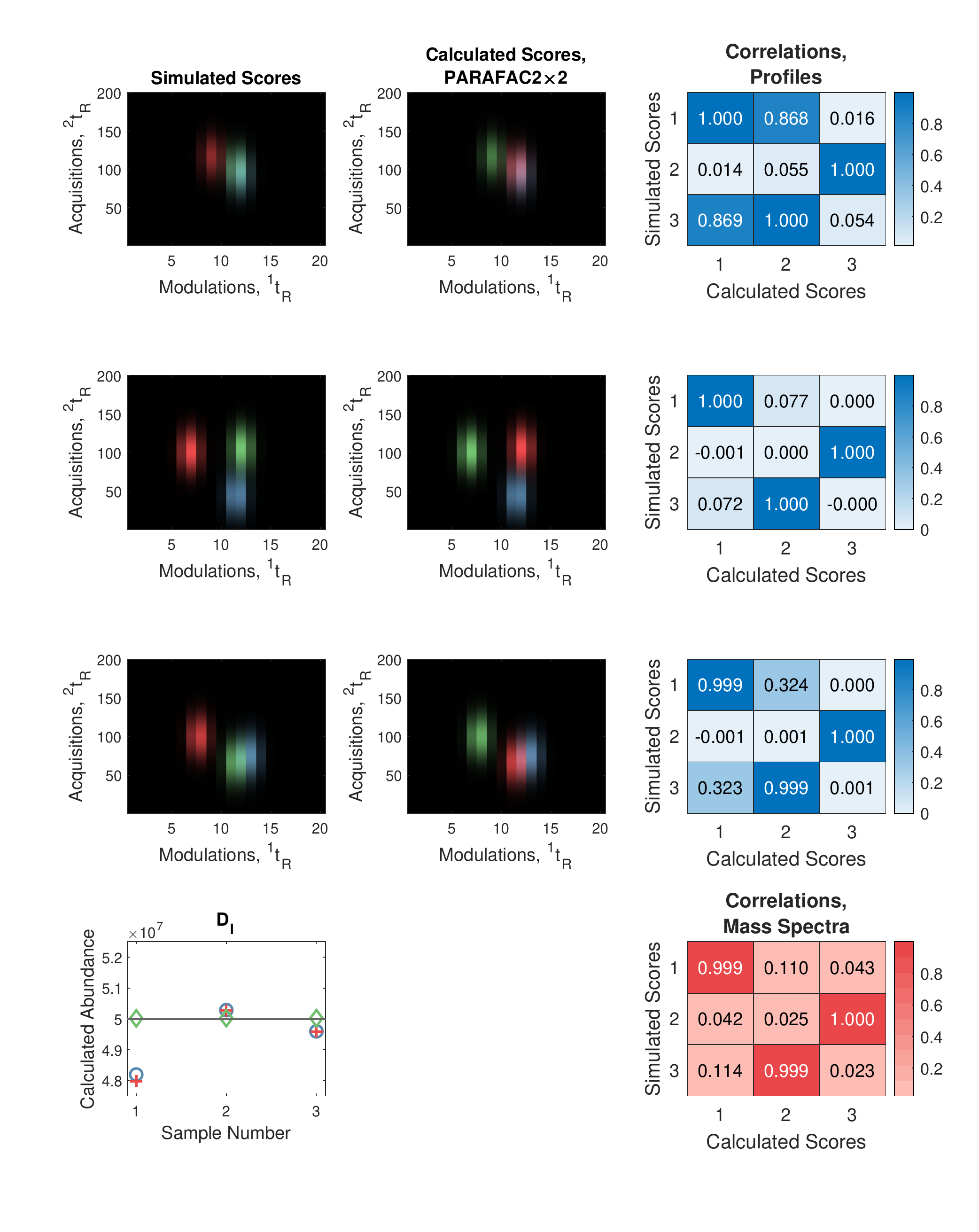}
    \caption{A simulated three-component model with a nominal SNR of 500. The percent variance explained using this model was 99.9565\%. These results are good, despite almost complete overlap between two components in sample 1.}
    \label{fig:my_label}
\end{figure}

The components in the synthetic data are indexed in different positions relative to the calculated features, but they can be related to each other using a permutation matrix, so this difference is inconsequential.

\subsection{Analysis of Calibration Data using PARAFAC$2\times2$}

Calibration data from a metabolomics study were used to test the PARAFAC2$\times$2 algorithm. Presumably, calibration data follows a predictable trend that can be used to judge the utility of PARAFAC2$\times$2 for quantitative and targeted analyses. 

A region of interest was excised from a calibration experiment, containing 67 different calibrants that were dissolved in an amenable organic solvent mixture (either 50\% Acetonitrile - 50\% Water, or 50\% Isopropanol - 50\% Toluene for polar and non-polar compounds respectively). Standard solutions were aliquotted at different volumes into 2-mL GC vials and blown down under nitrogen at 40 $^{\circ}$C. Residual moisture was removed by adding 100 $\mu$L of toluene dried under anhydrous sodium sulfate, which was again blown down using a stream of nitrogen at the same temperature. The dry residual extract was derivatized following a standard two-step methoximation / silylation approach. Briefly, 50 $\mu$L of 20 mg/mL methoxyamine HCl in pyridine at 60$^{\circ}$C for 2 hours, followed by 100 $\mu$L of MSTFA at 60$^{\circ}$ C for 1 hour. Based on the expected concentration of each standard within the pyridine/MSTFA solvent, and the 1-$\mu$L splitless injection volume, the relative quantification results from the PARAFAC2$\times$2 were plotted against the pg of analyte injected into the \gcxgc instrument.

The two analytes present in this region of interest are the trimethylsilyl (TMS) derivatives of salicylic acid and adipic acid (in this case, both derivatized molecules contained two TMS groups). Their identities were confirmed by examining their retention indices, mass spectra, and analyzing samples each containing a small fraction of analytes for confirmatory purposes.
\bigskip

\begin{figure}
    \begin{subfigure}{.5\textwidth}
    \centering
    \includegraphics[width = 0.9\linewidth]{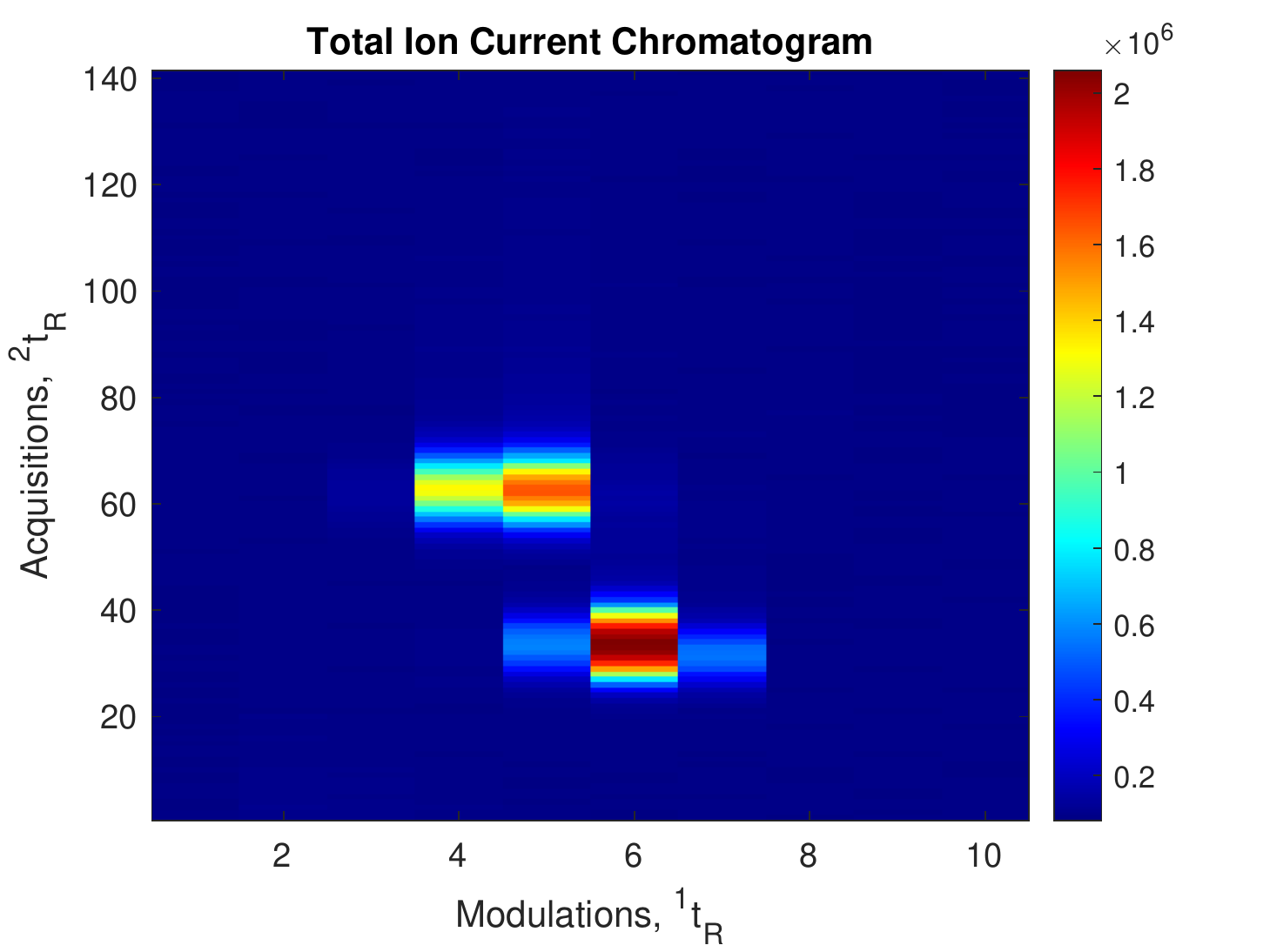}
    \caption{Raw chromatographic data of a representative sample: Total Ion Current (TIC)}
    \end{subfigure}
    \begin{subfigure}{.5\textwidth}
    \centering
    \includegraphics[width = .9\linewidth]{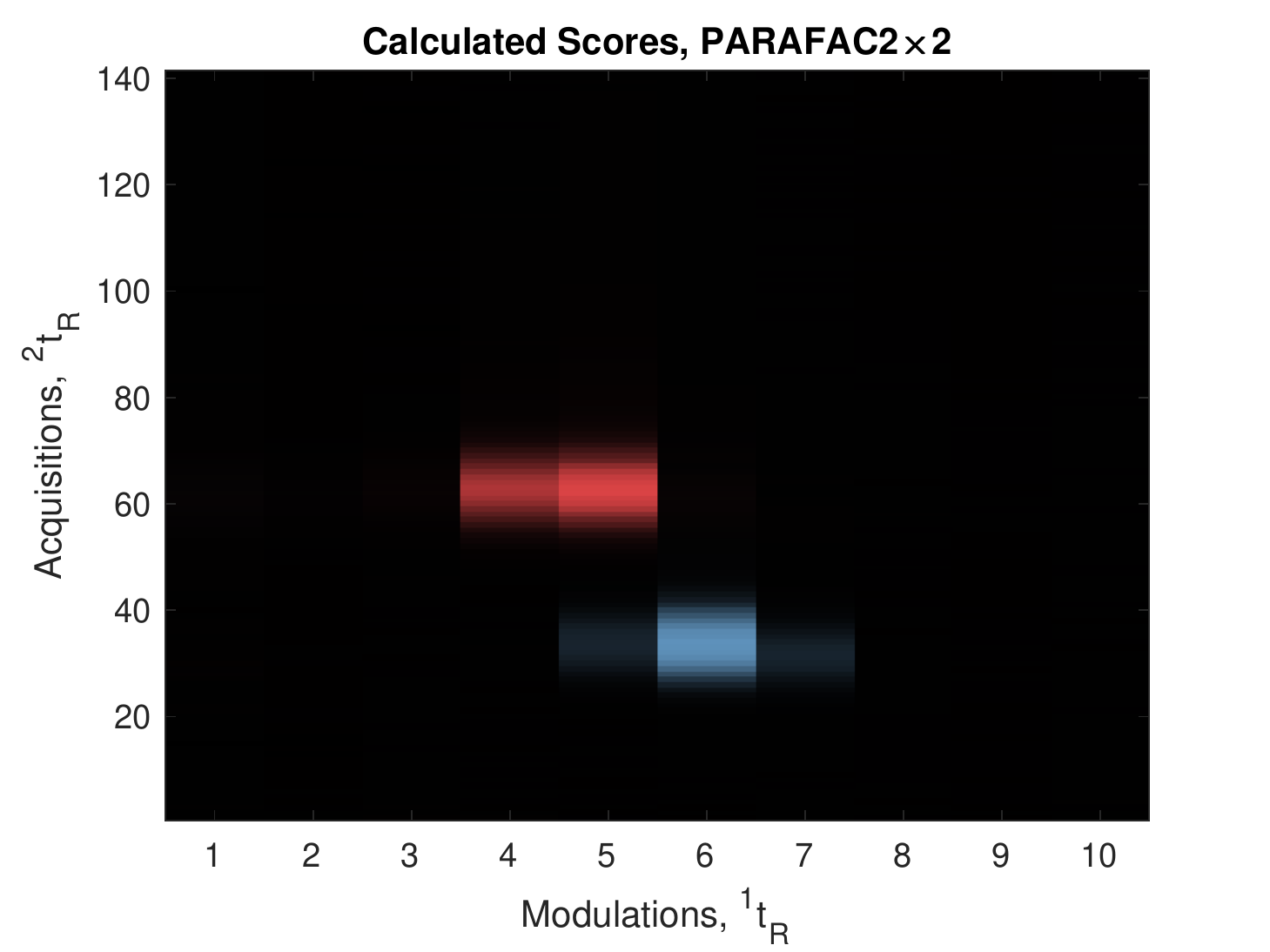}
    \caption{Extracted scores for a representative sample. The upper left component is adipic acid (2TMS), and the lower right component is salicylic acid (2TMS)}
    \end{subfigure}
    \newline
    \begin{subfigure}{.5\textwidth}
    \centering
    \includegraphics[width = 0.9\linewidth]{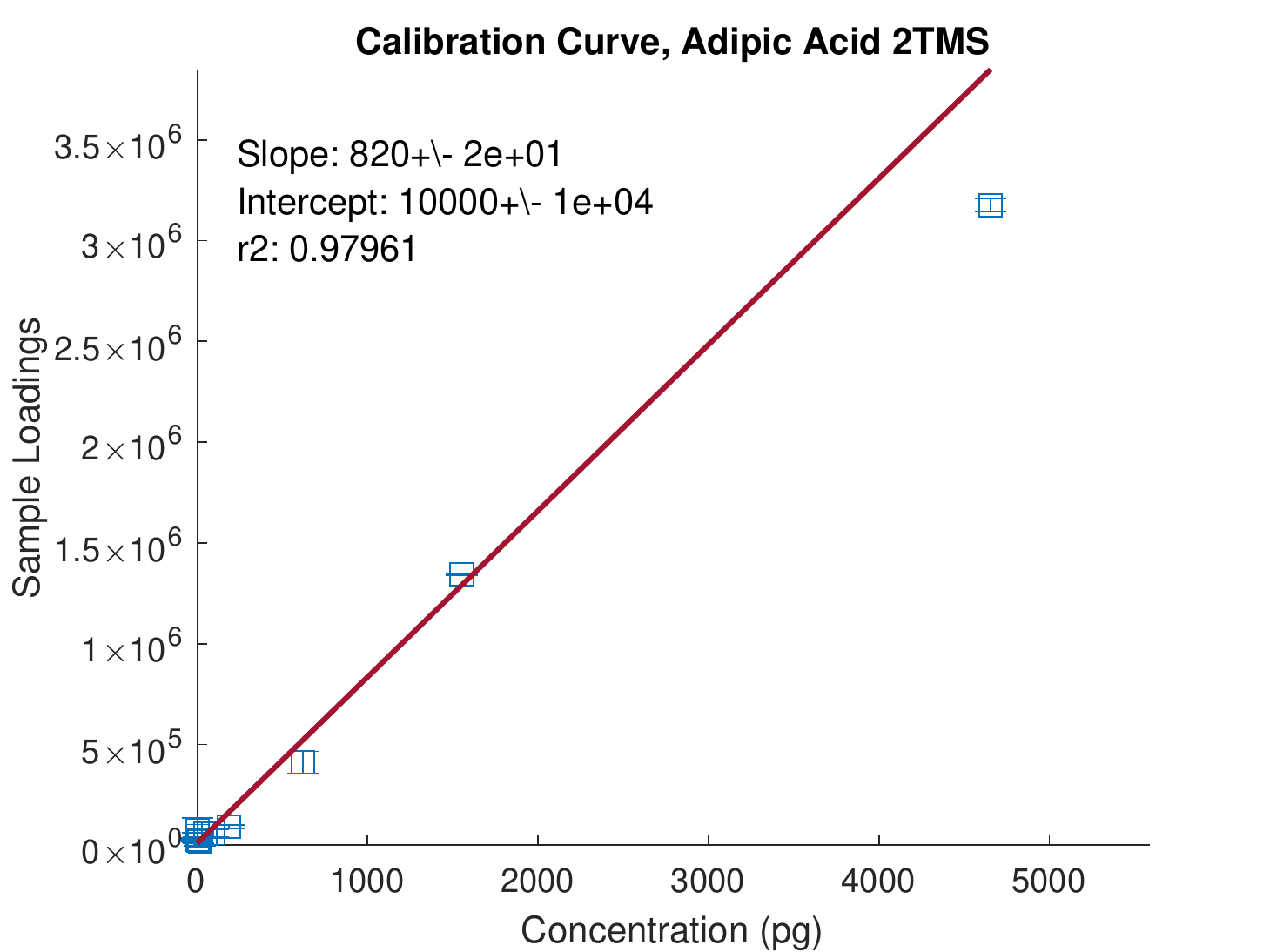}
    \caption{Calibration curve for adipic acid (2TMS)}
    \end{subfigure}
    \begin{subfigure}{0.5\textwidth}
    \centering
    \includegraphics[width = .9\linewidth]{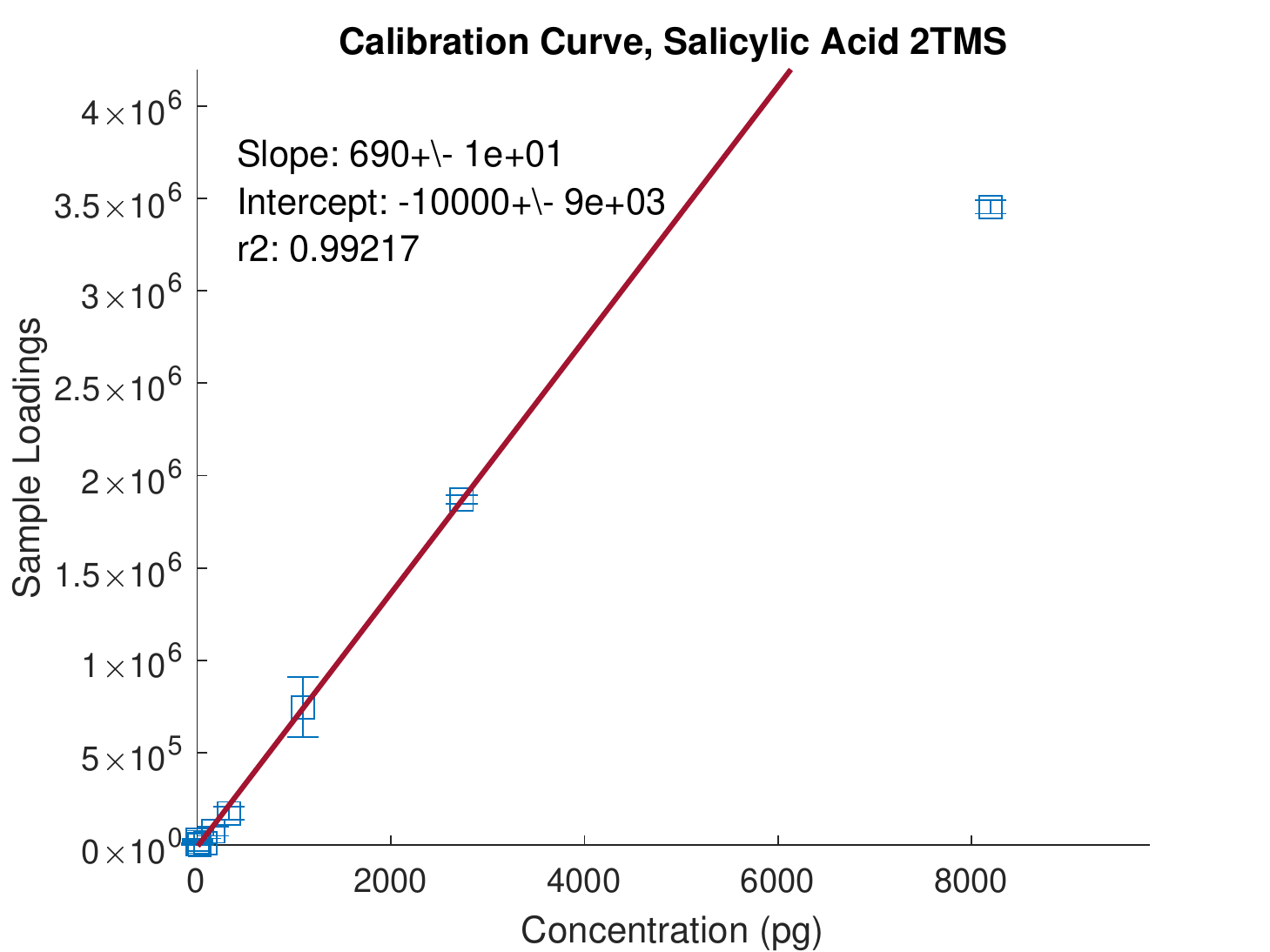}
    \caption{Calibration curve for salicylic acid (2TMS)}
    \end{subfigure}
    \newline
    \begin{subfigure}{0.5\textwidth}
    \centering
    \includegraphics[width = .9\linewidth]{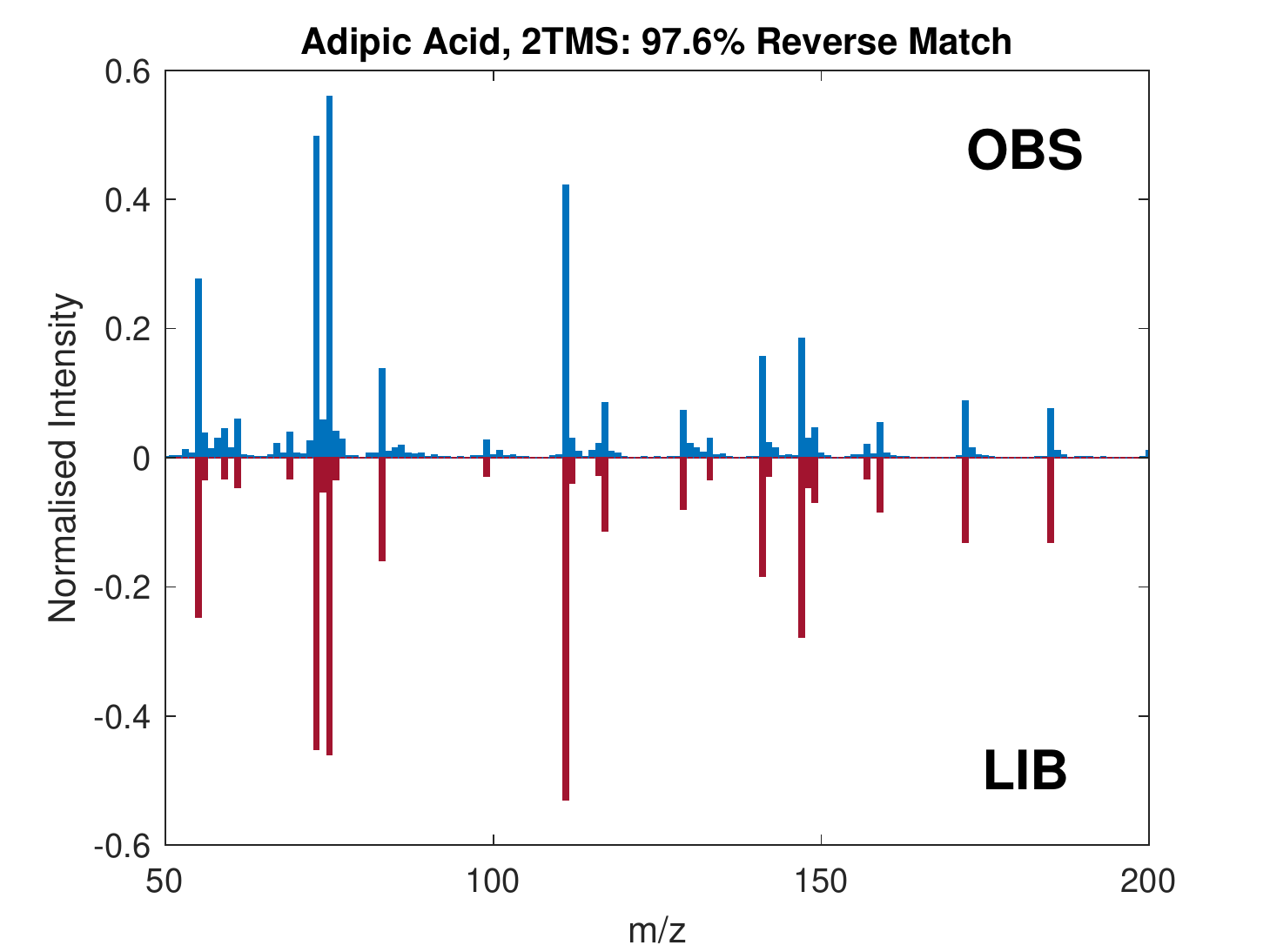}
    \caption{Library search results of the extracted mass spectrum for adipic acid (top), versus the library mass spectrum downloaded from the Golm Metabolome Database \cite{adms}}
    \end{subfigure}
    \begin{subfigure}{0.5\textwidth}
    \centering
    \includegraphics[width = .9\linewidth]{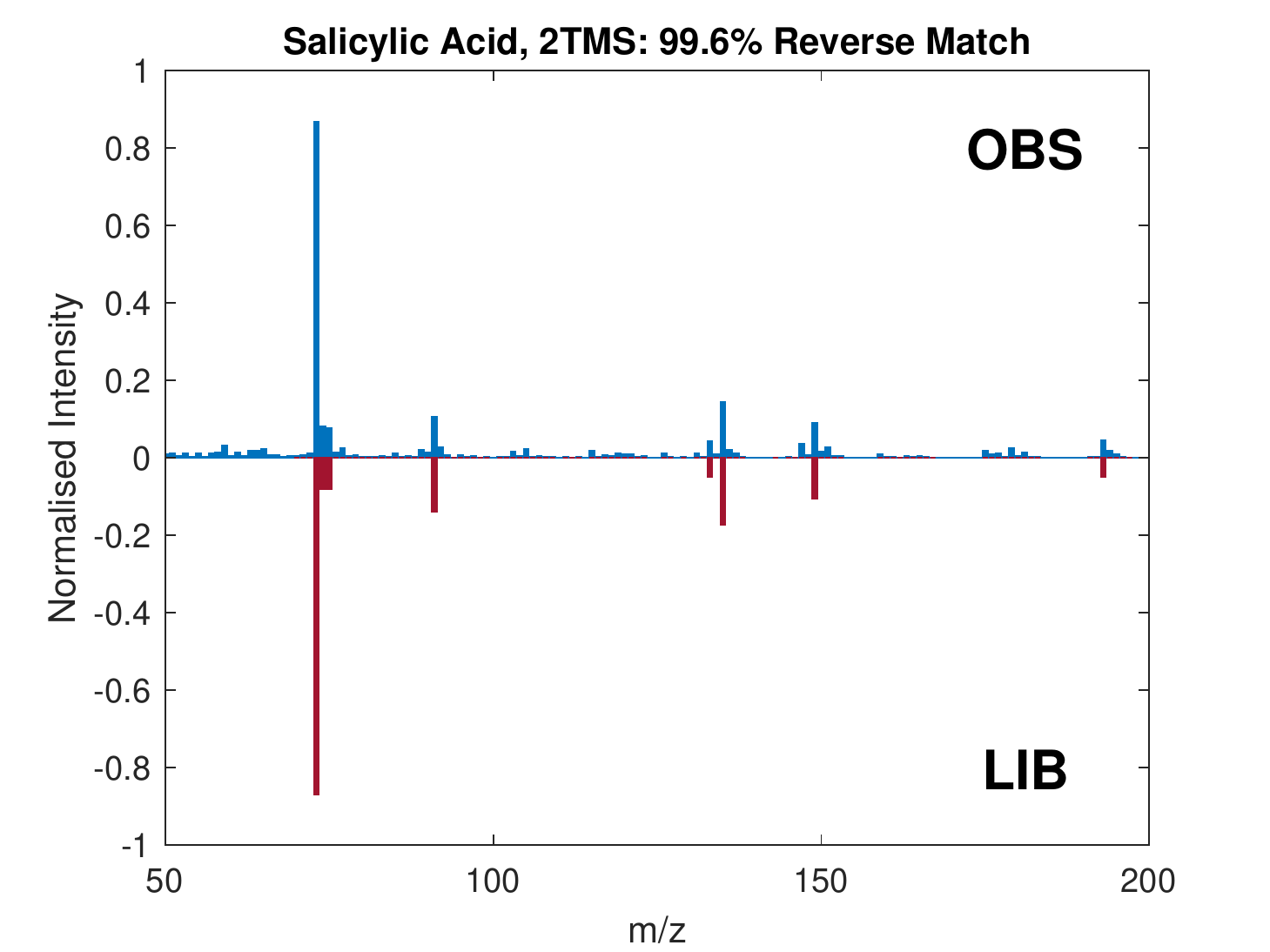}
    \caption{Library search results of the extracted mass spectrum for salicylic acid (top), versus the library mass spectrum downloaded from the Golm Metabolome Database \cite{salms}}
    \end{subfigure}
    \caption{Results of the analysis of calibration standards using PARAFAC2$\times$2. The analysis utilised three-factors, but the noise component was not displayed for ease of visualisation. Plots for the scores and TIC chromatograms are displayed for the 8\textsuperscript{th} sample in the calibration, where the calculated mass of the analyte on column was 2732.1 pg for salicylic acid (2TMS) and 1551.9 pg for adipic acid (2TMS)}

\end{figure}

\section{Extension to Multidimensional Separations Data}
Higher-order separations present an exciting new avenue of research for the analysis of complex samples. However, while it is not impossible to model \gcxgc data by unfolding the retention times as a single retention mode, unfolding scales poorly for higher-order chromatographic separations (e.g. GC$\times$ GC$\times $GC, LC$\times$LC$\times$LC, LC$\times$ GC$\times$GC, etc). In addition to the excess degrees of freedom, there are practical issues for calculating excessively large matrices, related to the available memory on the computer system being used for the calculations. Utilising higher-order separations has found more favour in the relatively new field of multidimensional liquid chromatography, since the peak widths are generally much larger and there are fewer practical limitations with regard to the sampling rate of the mass spectrometer \cite{venter2018comprehensive}. Some work has been done using comprehensive three-dimensional gas chromatography-time-of-flight mass spectrometry\cite{watson2017comprehensive}, but some issues persist with the published setup - since the instrumental sampling rate was limited to 200 Hz for third-dimension peaks eluting with a peak width of ~50 ms, the sampling rate for a single peak is limited to about 3-4 acquisitions per peak.

Consider a comprehensive three-dimensional separation, which can be described using an intuitive extension of Equation \ref{modes} for a 5\textsuperscript{th} order tensor, $\mathcal{X}_{ijklm} \in \mathbb{R}^{I\times J\times K\times L\times M}$ structured so that it contains $I$ acquisitions along the third retention mode, $K$ modulations from the second to the third dimension, and $L$ modulations from the first to the second separation dimension. $D_M$ represents the quantitative loadings for the $M^{th}$ sample, and $A$ is the matrix of $J\times R$ mass spectra.

Assuming that the practical aspects of higher-order separations with hyphenated multivariate detection methods (such as mass-spectrometers or spectroscopic methods) are overcome, it is possible to model $N$-dimensional drift using the same principles that guided the expressions developed previously for \gcxgc data, where $X \in \mathbb{R}^{I\times J*K*L*M}$

\begin{equation}\label{eq:parafac51}
    X = F_3(D_M\odot F_2\odot F_1\odot A)^T
\end{equation}

Or using unfolded data, where $X_m \in \mathbb{R}^{I*K*L \times J \times M}$:

\begin{equation}\label{eq:3dsep}
    X_m = (F_3\odot F_2\odot F_1)D_MA^T
\end{equation}

A model similar to the one proposed for \gcxgc data can be constructed for an $X_{ijklm}$ tensor with drift in three modes.

\begin{multline}
\mathcal{X}_{ijklm} = argmin(||X_{klm} - B_{klm}D_{klm}A_{klm}^T||^2_F+ \mu_{klm}||B_{klm} - P_{klm}B_{klm}^{*T}||^2_F \\ + ||X_{ilm} - B_{ilm}D_{ilm}A_{ilm}^T||^2_F + \mu_{ilm}||B_{ilm} - P_{ilm}B_{ilm}^{*T}||^2_F \\ + ||X_{ikm} - B_{ikm}D_{ikm}A_{ikm}^T||^2_F  + \mu_{ikm}||B_{ikm} - P_{ikm}B_{ikm}^{*T}||^2_F \\+ \mu_A||A_{klm} - A_{ilm}||_F^2 + \mu_A||A_{klm} - A_{ikm}||_F^2 + \mu_A||A_{ilm} - A_{ikm}||_F^2 )
\end{multline}

Where $X_{klm} \in \mathbb{R}^{I\times J\times K\times L\times M} = \mathbb{X}_{:,:,k,l,m}$, and other 3\textsuperscript{rd} order $X$ tensors follow similar convention, with the subscripts indexing what slices are being considered. Matrices associated with the non-negative PARAFAC2 decomposition are denoted by similar subscripts.

The number of terms that restrict the dissimilarity of the mass spectra with respect to the different methods of unfolding that data are related to the number of possible combinations of each of the terms, which is equal to the binomial coefficient, $_NC_2$. For even higher orders of separations further extensions are possible, but it is not convenient nor especially useful to come up with a generalised notation for these circumstances. The authors leave this exercise to the interested reader.

\section{Conclusions}
A general theory of modelling separations data with drift in multiple modes is proposed, and has been shown to work on experimental and synthetic data that are close to the worst possible scenario for independent chromatographic drift in two modes. The presents a parsimonious method for the deconvolution of signals and extraction of both qualitative and quantitative metrics from \gcxgc data, and eliminates the need for dynamic programming routines that may contribute to peak splitting and/or peak drop-out commonly encountered in \gcxgc peak tables.

For targeted analysis of \gcxgc data, this algorithm is sufficient. Determining appropriate regions of interest and a value for the component number is a relatively simple task for a handful of components. Since the component number is specific to each region, the number of parameters required scales with the number of components being analysed in a series of chromatograms. While the number of data analysis parameters does not scale with the number of components being analysed using the currently available commercial offerings, there is still certainly a high degree of complexity inherent to analysing entire chromatograms using a single set of parameters, and there is no guarantee that a single set of parameters will be sufficient to analyse all of the desired targets with a high degree of accuracy. The approach proposed by the authors is more flexible, similar to the application of PARAFAC2 to GC-MS experiments, but requires skilled user intervention. This is a significant first step towards a holistic chemometric method for pre-processing entire \gcxgc experiments, but additional work is required to automate the selection of regions of interest and choosing appropriate component numbers for each region.

It is also possible to model single samples using PARAFAC2$\times$2, since a high number of replicates are inherent even with a single \gcxgc sample. This makes modelling of single chromatograms extensible to larger numbers of chromatograms using the same model, and the results of the analysis for one sample may be extrapolated to several more.

Additional investigations are needed to evaluate this technique in relation to different methods for unfolding the data, which is not a trivial task, and is deserving of its own article. Considerations for the practicality of unfolding data in a way that generates more replicates at the expense of degrees of freedom, and computational efficiency must be considered. However, to the best of the authors' knowledge this algorithm is the first of its kind applied to multidimensional chromatographic data, and represents a significant leap forward for the field of \gcxgc data analysis.

\section{Acknowledgements}

The authors would like to thank Rasmus Bro for a number of productive discussions.
%However in this case, since $P_k$ may vary for each sample a single pair of matrices matrix cannot be used to describe chromatographic drift across multiple samples using the KRB product as in Equation \ref{parafackrb}. Between $L$ samples, for an $i\times j\times k$ region of the chromatogram, chemical components may drift independently of one another between chromatographic runs due to variations in the operating conditions of the instrument that do not affect different chemical components in equal measure. The goal for modelling \gcxgc data is to solve for individual elution profiles, which can be thought of as a series of parallel vectors unfolded to allow for drift along a single mode, or more intuitively as am $I \times R\times K$ tensor of two dimensional scores:

%\begin{equation}
%    \mathcal{X}_{I\times J\times K\times L} = \mathcal{F}_{I\times R\times K\times L} \mathcal{D}_{R\times R\times K\times L}A^T_{J\times R}
%\end{equation}

\appendix

\section{Khatri-Rao Product}

The KR product is commonly used as the tensor product, owing to the simplicity by which the PARAFAC model can be optimised using the Alternating Least Squares (ALS) algorithm that is analogous to the way in which bilinear models are traditionally optimised. 

The KR product is defined as the column-wise Kronecker product for two matrices, $A$ and $B$ with an equal number of columns, $R$, that correspond to the number of chemical factors:

\begin{equation}\label{app:KR}
    A = [a_1, a_2, ..., a_R] 
\end{equation}

\begin{equation}
    B = [b_1, b_2, ..., b_R]
\end{equation}

\begin{equation}
    A\odot B = [A_1 \otimes B_1, A_2 \otimes B_2, ..., A_R \otimes B_R]
\end{equation}

\section{Derivation of an Expression to Solve for $A_{kl}$, and $A_{il}$}

An expression that minimises the sum of squared residuals can also be described as the minimisation of the square of the Frobenius Norm:

$$
||A||_F = \sqrt{\sum_{i=1}^m\sum_{j=1}^n|a_{ij}|^2} = \sqrt{tr(A^TA)} = \sqrt{tr(AA^T)}
$$

An expression for an estimate of $A_{kl}$ or $A_{il}$ takes into account the coupling term that controls the difference between the two expressions relative to the mass spectral coupling constant, $\mu_A$. Deriving an expression for $A_{kl}$ begins with calculating the derivative of the following expression with respect to $A_{kl}$:

$$
\frac{\partial}{\partial A_{kl}}(||X_{kl} - B_{kl}D_{kl}A^T||^2_F + \mu_A||A_{kl} - A_{il}||_F^2) = 0
$$

Which has been simplified from Equation \ref{fullexp}, since the derivative with respect to $A_{kl}$ of $||X_{il} - B_{il}D_{il}A^T||^2_F + \mu_{il}||B_{il}-P_{il}B^*||^2_F$ and $\mu_{kl}||B_{kl}-P_{kl}B^*||^2_F$ are both zero. Expanding the terms in the previous expression and adding an arbitrary constant, $\frac{1}{2}$, to aid in simplification:

$$
\frac{\partial}{\partial A_{kl}} \frac{1}{2}tr\left( (X_{kl} - B_{kl}D_{kl}A_{kl}^T)^T(X_{kl}-B_{kl}D_{kl}A_{kl}^T)\right) + \frac{\partial}{\partial A_{kl}}tr\left(\mu_A(A_{kl} - A_{il})^T(A_{kl} - A_{il})\right) = 0
$$

\begin{multline*}
\frac{\partial}{\partial A_{kl}} tr(\left( X_{kl}^TX_{kl}\right) - \frac{\partial}{\partial A_{kl}}tr \left(X_{kl}^TB_{kl}D_{kl}A_{kl}^T\right)\\ - \frac{\partial}{\partial A_{kl}}tr\left(A_{kl}D_{kl}B_{kl}^TX_{kl}\right) + \frac{\partial}{\partial A_{kl}}tr\left(A_{kl}D_{kl}B_{kl}^TB_{kl}D_{kl}A_{kl}^T\right)\\ + \frac{\partial}{\partial A_{kl}}tr\left(\mu_A\left(A_{kl}^TA_{kl} - A_{kl}^TA_{il} - A_{il}^TA_{kl} + A_{il}^TA_{il}\right)\right) = 0
\end{multline*}

Note that the term $\frac{\partial}{\partial A}\left(A_{kl}^TA_{kl} - A_{kl}^TA_{il}^T - A_{il}^TA_{kl} + A_{il}^TA_{il}\right)$ with respect to $A_{kl}$ is equivalent when the derivative is taken with respect to $A_{il}$. This reveals that the expression for the estimation of $A_{kl}$ is the same expression as the expression that estimates $A_{il}$.

Using the following identities from the Matrix Cookbook:

$$
\frac{\partial}{\partial A_{kl}}tr \left(X_{kl}^TX_{kl}\right) = 0
$$

$$
\frac{\partial}{\partial A_{kl}}tr \left(X_{kl}^TB_{kl}D_{kl}A_{kl}^T\right) = X_{kl}^TB_{kl}D_{kl}
$$

$$
\frac{\partial}{\partial A_{kl}}tr\left(A_{kl}D_{kl}B_{kl}^TX_{kl}\right) = X_{kl}^TB_{kl}D_{kl}
$$

$$
\frac{\partial}{\partial A_{kl}}tr\left(A_{kl}D_{kl}B_{kl}^TB_{kl}D_{kl}A_{kl}^T\right) = 2A_{kl}D_{kl}B_{kl}^TB_{kl}D_{kl}
$$

$$
\frac{\partial}{\partial A_{kl}}tr\left(A_{kl}^TA_{kl}\right) = 2A_{kl}
$$

$$
\frac{\partial}{\partial A_{kl}}tr\left(A_{kl}^TA_{il}\right) = A_{il}
$$

$$
\frac{\partial}{\partial A_{kl}}tr\left(A_{il}^TA_{kl}\right) = A_{il}
$$

$$
\frac{\partial}{\partial A_{kl}}tr\left(A_{il}^TA_{il}\right) = 0
$$

Substituting into the previous expression yields:

$$
-2X_{kl}^TB_{kl}D_{kl} + 2A_{kl}D_{kl}B_{kl}^TB_{kl}D_{kl} +2\mu_AA_{kl} - 2\mu_AA_{il} = 0
$$

$$
2A_{kl}D_{kl}B_{kl}^TB_{kl}D_{kl} +2\mu_AA_{kl} = 2X_{kl}^TB_{kl}D_{kl} +  2\mu_AA_{il}
$$

Solving for $A_{kl}$

$$
A_{kl} = \frac{\mu_AA_{il} + X_{kl}^TB_{kl}D_{kl}}{D_{kl}B_{kl}^TB_{kl}D_{kl} + \mu_AI_R}
$$

\section{Derivation of an Expression to Solve for $B_{k}$,$B_{kl}$, and $B_{il}$}

An expression for $B_{kl}$ and $B_{il}$ can be solved for with respect to their respective coupled terms, and only differ with respect to the arrangement of the data, and is agnostic to the mass spectral coupling term, $\mu_A$. As such, the following derivation can be described for the flexible coupling method of calculating non-negative PARAFAC2 with respect to $B_k$. This solution was previously published by Michael Armstrong on Mathematics StackExchange \cite{3994699}.

$$
\frac{\partial}{\partial B_{k}}(||X_k - B_kD_kA^T||^2 + \mu_k||B_k-P_kB^*||^2) = 0
$$

We can rearrange the original equation, adding an arbitrary constant as before, $\frac{1}{2}$, to aid with simplification later on.

$$
\frac{\partial}{\partial B_{k}}\frac{1}{2}tr((X_k - B_kD_kA^T)(X_k - B_kD_kA^T)^T) + \frac{\partial}{\partial B_{k}}\frac{1}{2}\mu_k(tr((B_k - P_kB^*)(B_k-P_kB^*)^T) = 0
$$

Expanding the equations, where $(X_k - B_kD_kA^T)^T = X_k^T - AD_kB_k^T$ and $(B_k - P_kB^*)^T = B_k^T - B^{*T}P_k^T$:

\begin{multline*}
\frac{\partial}{\partial B_k}\frac{1}{2}(tr(X_kX_k^T) -tr(X_kAD_kB_k^T) - tr(B_kD_kA^TX_k^T) + tr(B_kD_kA^TAD_kB_k^T))\\ +\frac{\partial}{\partial B_k}\frac{1}{2}\mu_k(tr(B_kB_k^T)-tr(B_kB^{*T}P_k^T) - tr(P_kB^*B_k^T) + tr(P_kB^*B^{*T}P_k^T)) = 0
\end{multline*}

This equation can be simplified by using some convenient identities from [the Matrix Cookbook](https://www.math.uwaterloo.ca/~hwolkowi/matrixcookbook.pdf):
$$
\frac{\partial}{\partial{B_k}}tr(X_kX_k^T) = 0
$$

$$
\frac{\partial}{\partial{B_k}}tr(X_kAD_kB_k^T) = X_kAD_k
$$

$$
\frac{\partial}{\partial{B_k}}tr(B_kD_kA^T) = X_kAD_k
$$

$$
\frac{\partial}{\partial{B_k}}tr(B_kD_kA^T) = X_kAD_k
$$

$$
\frac{\partial}{\partial{B_k}}tr(B_kD_kA^TAD_kB_k^T) = 2B_kD_kA^TAD_k
$$

$$
\frac{\partial}{\partial{B_k}}tr(B_kB^{*T}P_k^T) = PkB^*
$$

$$
\frac{\partial}{\partial{B_k}}tr(P_kB^*B_k^T) = P_kB^*
$$

$$
\frac{\partial}{\partial{B_k}}tr(P_kB^*B^{*T}P_k^T) = 0
$$

$$
\frac{\partial}{\partial{B_k}}tr(B_kB_k^T) = 2B_k
$$

Substituting into the previous equation yields:

$$
-X_kAD_k + B_kD_kA^TAD_k + \mu_kB_k - \mu_kP_kB^* = 0
$$

Multiplying by the inverse of $D_kA^TAD_k$:

$$
-X_kAD_k(D_kA^TAD_k)^{-1} + B_k + \mu_kB_k(D_kA^TAD_k)^{-1} - \mu_kP_kB^*(D_kA^TAD_k)^{-1} = 0
$$

Solving for $B_k$:

$$
-X_kAD_k + \mu_kB_k - \mu_kP_kB^* = -B_k(D_kA^TAD_k)
$$

$$
-X_kAD_k - \mu_kP_kB^* = -B_k(D_kA^TAD_k) - \mu_kB_k
$$

Yields the final form of the equation:
$$
B_k = \frac{X_kAD_k + \mu_kP_kB^*}{D_kA^TAD_k + \mu_kI_R}
$$

This solution is similar to the one released in Jeremy Cohen's software package \cite{nnparafac2m}, although the authors could not a previously published derivation.

 \bibliographystyle{elsarticle-num} 
 \bibliography{cas-refs}





% \end{thebibliography}
\end{document}